\documentstyle[preprint,aps]{revtex}
\draft

\newcommand{\ba}{\begin{eqnarray}}
\newcommand{\ea}{\end{eqnarray}}
\begin{document}
\pagestyle{plain}

\title{Strong decays of nonstrange $q^3$ baryons}
\author{R. Bijker$^{a}$, F. Iachello$^{b}$ and A. Leviatan$^{c}$}
\address{$^{a}$Instituto de Ciencias Nucleares, U.N.A.M.,
A.P. 70-543, 04510 M\'exico, D.F., M\'exico
\newline
$^{b}$Center for Theoretical Physics, Sloane Laboratory,
Yale University,\\
New Haven, CT 06520-8120, U.S.A.
\newline
$^{c}$Racah Institute of Physics, The Hebrew University,
Jerusalem 91904, Israel}

\maketitle

\begin{abstract}
We study strong decays of nonstrange baryons by making use of the algebraic
approach to hadron structure. Within this framework we derive closed
expressions for decay widths in an elementary-meson emission model and
use these to analyze the experimental data for
$N^{\ast} \rightarrow N + \pi$,
$N^{\ast} \rightarrow \Delta + \pi$,
$N^{\ast} \rightarrow N + \eta$,
$\Delta^{\ast} \rightarrow N + \pi$,
$\Delta^{\ast} \rightarrow \Delta + \pi$ and
$\Delta^{\ast} \rightarrow \Delta + \eta$ decays.
\end{abstract}

\begin{center}
PACS numbers: 13.30.Eg, 11.30.Na, 14.20.Gk
\end{center}

\newpage
\section{Introduction}
\setcounter{equation}{0}

A large amount of experimental data was accumulated in the 1960's
and 1970's on the spectroscopy of light hadrons. These data led first
to the introduction of $SU_f(3)$ by Gell'Mann \cite{MGM} and Ne'eman
\cite{YN} (later enlarged to $SU_{sf}(6)$ by G\"ursey and Radicati
\cite{GR}), and subsequently to $SU_c(3)$ color symmetry as a gauge
symmetry of strong interactions. The construction of
dedicated facilities ({\it e.g.} CEBAF, MAMI, ELSA) that promise to produce
new and more accurate data on the same subject, have stimulated us
to revisit baryon spectroscopy with the intent to reexamine whether
or not the new data can shed some new light on the structure of hadrons.
In this reanalysis we have introduced, in addition to the basic
spin-flavor-color symmetry, $SU_{sf}(6) \otimes SU_c(3)$, a new
ingredient, namely a space symmetry, ${\cal G}$, which we have taken to be
${\cal G}=U(7)$ for baryons \cite{BIL}. The introduction of the space
symmetry allows us to examine, in a straightforward way, several
limiting situations ({\it e.g.} harmonic oscillator and collective
dynamics), and to produce transparent results that can be
used to analyze the experimental data.
This approach has been used \cite{BIL,emff}
to analyze the mass spectrum and electromagnetic couplings of
nonstrange baryon resonances. It presents an alternative for
the use of nonrelativistic or relativized Schr\"odinger equations.
In addition to electromagnetic couplings, strong decays of baryons
provide an important, complementary, tool to study the structure of
baryons. These strong couplings are needed to analyze the new
upcoming experimental data, especially ($e,e^{\prime}\pi$) and
($e,e^{\prime}\eta$). Furthermore, we want
to understand whether or not unusual features appear in the data, which
may point out to `new' physics, new meaning here unconventional
configurations of quarks and gluons, such as hybrid states, $q^3-g$, or
multiquark states, $q^3-q \bar{q}$. For example, the observed
large $\eta$ width of the $N(1535)S_{11}$ resonance has led to
considerable discussion \cite{nimai,zli,Kaiser,LW,riska}
about the nature of this resonance.

The strong decays have been analyzed previously in the
nonrelativistic \cite{KI} and relativized quark models \cite{CR}.
These models emphasize single-particle aspects of quark dynamics
in which only a few low-lying configurations in the confining potential
contribute significantly to the baryon wave function. In the framework
of the earlier mentioned algebraic approach, it is possible to study
also other, more collective, types of dynamics. In this article,
we analyze in detail the strong decays in a collective model of
baryon structure.
The article is organized as follows: in Section~\ref{sec2} we briefly
discuss the method of calculation, in Section~\ref{sec3} we present the
results, which are compared with the existing data in Section~\ref{sec4}.
Finally, in Section~\ref{sec5} we present our conclusions and point out
some open problems.

\section{Method of calculation}
\label{sec2}

We consider in this article strong decays of baryons by the emission of
a pseudoscalar meson
\ba
B \rightarrow B^{\prime} + M ~.
\ea
In order to calculate these decays we need two ingredients: (i) the
wave function of the initial and final states and (ii) the form of the
transition operator. We write both the wave functions and the transition
operators in algebraic form. We take the wave functions as
representations of $U(7) \otimes SU_{sf}(6) \otimes SU_c(3)$, as
discussed in \cite{BIL}. The algebraic formulation allows us to study
a large class of models, all with the same spin-flavor-color structure,
but different types of quark dynamics ({\it e.g.} single-particle
and collective). Each scenario corresponds to
different ways in which the $U(7)$ spatial symmetry is broken
down to the angular momentum group $SO(3)$. Among the models in this
class is the familiar harmonic oscillator quark model
characterized by the breaking
\ba
U(7) \supset U(6) \supset U_{\rho}(3) \otimes U_{\lambda}(3)
\supset SO_{\rho}(3) \otimes SO_{\lambda}(3) \supset SO(3)
\supset SO(2) ~,
\ea
where the index $\rho$ and $\lambda$ refers to the two relative
Jacobi coordinates
\ba
\vec{\rho} &=& \frac{1}{\sqrt{2}}
(\vec{r}_1 - \vec{r}_2) ~,
\nonumber\\
\vec{\lambda} &=& \frac{1}{\sqrt{6}}
(\vec{r}_1 + \vec{r}_2 - 2\vec{r}_3) ~. \label{jacobi}
\ea
The nonrelativistic and relativized quark potential models
\cite{IK,CI} are formulated in this harmonic oscillator basis.

In a collective model the baryon resonances
are interpreted in terms of rotations and vibrations of the
string configuration in Fig.~\ref{geometry}, which is characterized
by the two relative Jacobi coordinates of Eq.~(\ref{jacobi}).
The three constituent parts carry the global quantum numbers:
$\mbox{flavor}=\mbox{triplet}=u,d,s$;
$\mbox{spin}=\mbox{doublet}=1/2$; and $\mbox{color}=\mbox{triplet}$
(we do not consider here heavy quark flavors).
Different types of collective models are specified by a
distribution of the charge, magnetic moment, etc.,
over the entire volume. For the present analysis of strong
decays we use the (normalized) distribution
\ba
g(\beta) &=& \beta^2 \, \mbox{e}^{-\beta/a}/2a^3 ~, \label{gbeta}
\ea
where $\beta$ is a radial coordinate and $a$ is a scale parameter.
We have shown in \cite{emff} that the above distribution
appears to describe the available data on electromagnetic form
factors and helicity amplitudes up to $Q^2 \approx 10-20$ (GeV/c)$^2$.

The spatial part of the baryon wave functions in the `collective' model
is characterized by
the following labels: $(v_1,v_2);K,L^P_t$ \cite{BIL}, where $(v_1,v_2)$
denotes the vibrations (stretching and bending) of the configuration
of Fig.~\ref{geometry}; $K$ denotes the projection of the rotational
angular momentum $L$ on the body-fixed axis; $P$ the parity and $t$
the symmetry type of the state under the point group $D_3$.
The classification under $D_3$ and parity is equivalent to that of the
point group $D_{3h}$, which describes the discrete symmetry of the
object of Fig.~\ref{geometry}.
Instead of the $D_3$ labels ($t=A_1$, $A_2$, $E$)
one can use the labels of $S_3$ ($t=S$, $A$, $M$),
the group of permutations of the three constituent parts
($S_3$ and $D_3$ are isomorphic).
The permutation symmetry $t$ of the spatial part of the
wave function must be the same as that of the spin-flavor part
in order to have total wave functions that are antisymmetric (the color
part is a color-singlet, {\it i.e.} antisymmetric). Therefore one
can also use the dimension of the $SU_{sf}(6)$ representations to label
the states:
$A_1 \leftrightarrow S \leftrightarrow 56$,
$A_2 \leftrightarrow A \leftrightarrow 20$ and
$E   \leftrightarrow M \leftrightarrow 70$.
We adhere to the latter notation and label the states by
\ba
\left| \, [\mbox{dim}\{SU_{sf}(6)\},L^P]_{(v_1,v_2);K} \, \right> ~.
\ea
When the spin-flavor quantum numbers are explicitly added, we obtain
the total baryon wave function
\ba
\left| \, ^{2S+1}\mbox{dim}\{SU_f(3)\}_J \,
[\mbox{dim}\{SU_{sf}(6)\},L^P]_{(v_1,v_2);K} \, \right> ~, \label{wf}
\ea
where $S$ and $J$ are the spin and total angular momentum
$\vec{J}=\vec{L}+\vec{S}$~.
For example, in this notation the nucleon and delta wave functions
are given by
\ba
\left| \, ^{2}8_{1/2} \, [56,0^+]_{(0,0);0} \, \right>
\hspace{1cm} \mbox{and} \hspace{1cm}
\left| \, ^{4}10_{3/2} \, [56,0^+]_{(0,0);0} \, \right> ~,
\ea
respectively.

The second ingredient in the calculation is the form of the operator
inducing the strong transition. Several forms have been suggested
\cite{LeYaouanc}. We use here the simple form \cite{KI}
\ba
{\cal H} &=& \frac{1}{(2\pi)^{3/2} (2k_0)^{1/2}}
\sum_{j=1}^{3} X^M_{j} \left[
2g \, (\vec{s}_j \cdot \vec{k}) \mbox{e}^{-i \vec{k} \cdot \vec{r}_j}
+ h \, \vec{s}_j \cdot
(\vec{p}_j \, \mbox{e}^{-i \vec{k} \cdot \vec{r}_j} +
\mbox{e}^{-i \vec{k} \cdot \vec{r}_j} \, \vec{p}_j) \right] ~, \label{hs}
\ea
where $\vec{r}_j$, $\vec{p}_j$ and $\vec{s}_j$ are the coordinate,
momentum and spin of the $j$-th constituent, respectively;
$k_0=E_M=E_B-E_{B^{\prime}}$ is the meson energy, and
$\vec{k}=\vec{P}_M=\vec{P}-\vec{P}^{\prime}=k \hat z$ denotes the momentum
carried by the outgoing meson. Here $\vec{P}=P_z \hat z$
and $\vec{P}^{\prime}$ ($=P^{\prime}_z \hat z$)
are the momenta of the initial and final baryon.
The coefficients $g$ and $h$ denote the strength of the two terms
in the transition operator of Eq.~(\ref{hs}).
The flavor operator $X^M_j$ (to be discussed below)
corresponds to the emission of an
elementary meson ($M$) by the $j$-th constituent:
$q_j \rightarrow q_j^{\prime} + M$ (see Figure~\ref{qqM}).

Using the symmetry of the wave functions for nonstrange baryons
(the only case we discuss here), transforming to Jacobi
coordinates and integrating over the baryon center of mass coordinate,
we find
\ba
{\cal H} &=& \frac{1}{(2\pi)^{3/2} (2k_0)^{1/2}} \,
6 X^M_{3} \left[ g \, k s_{3,z} \, \hat U
- h \, s_{3,z} (\hat T_z - \frac{1}{6}(P_z+P^{\prime}_z) \hat U)
- \frac{1}{2} h \, (s_{3,+} \hat T_- + s_{3,-} \hat T_+) \right] ~,
\nonumber\\
\label{hstrong}
\ea
with
\ba
\hat U &=& \mbox{e}^{i k \sqrt{\frac{2}{3}} \lambda_z} ~,
\nonumber\\
\hat T_m &=& \frac{1}{2} \left( \sqrt{\frac{2}{3}} \, p_{\lambda,m} \,
\mbox{e}^{i k \sqrt{\frac{2}{3}} \lambda_z} +
\mbox{e}^{i k \sqrt{\frac{2}{3}} \lambda_z} \,
\sqrt{\frac{2}{3}} \, p_{\lambda,m} \right) ~. \label{utop}
\ea
By using momentum conservation $P_z+P^{\prime}_z$ can be written as
$2P_z - k$. The dependence on the overall center of mass momentum is
spurious and must be eliminated. (In the rest frame of $B$, $P_z=0$
and the spurious term is automatically absent.)
By making the replacement,
$\vec{p}_{\lambda}/m_q \rightarrow -i k_0 \vec{\lambda}$ \cite{MB},
and the mapping onto the algebraic operators,
$\sqrt{2/3} \, \lambda_m \rightarrow \beta \hat D_{\lambda,m}/X_D$
\cite{BIL,emff}, we can write Eq.~(\ref{utop}) as
\ba
\hat U &=& \mbox{e}^{i k \beta \hat D_{\lambda,z}/X_D} ~,
\nonumber\\
\hat T_m &=& - \frac{i m_q k_0 \beta}{2X_D} \left(
\hat D_{\lambda,m} \, \mbox{e}^{i k \beta \hat D_{\lambda,z}/X_D} +
\mbox{e}^{i k \beta \hat D_{\lambda,z}/X_D} \,
\hat D_{\lambda,m} \right) ~.
\ea
The dipole operator $\hat D_{\lambda,m}$ is a generator of $U(7)$ and
$X_D$ is its normalization, as discussed in \cite{BIL,emff}.
The calculation of the matrix elements of ${\cal H}$
can be done in configuration space ($\vec{\rho}$, $\vec{\lambda}$)
or in momentum space ($\vec{p}_{\rho}$, $\vec{p}_{\lambda}$). The mapping
onto the algebraic space of $U(7)$ is a convenient way to carry out the
calculations, much in the same way as the mapping of coordinates and
momenta onto creation and annihilation operators in the harmonic
oscillator space.

In the collective model the matrix elements of ${\cal H}$ are obtained
by folding with the distribution function $g(\beta)$ of Eq.~(\ref{gbeta}).
These collective matrix elements can be expressed in terms of
helicity amplitudes. For decays in which the initial baryon has angular
momentum $\vec{J}=\vec{L}+\vec{S}$ and in which the final baryon is either
the nucleon or the delta with $L^{\prime}=0$ and thus
$J^{\prime}=S^{\prime}$, the helicity amplitudes are
\ba
A_{\nu}(k) &=& \int d \beta \, g(\beta) \, \langle \alpha^{\prime},
L^{\prime}=0,S^{\prime},J^{\prime}=S^{\prime},M_J^{\prime}=\nu \,
| \, {\cal H} \, | \, \alpha,L,S,J,M_J=\nu \rangle
\nonumber\\
&=&  \frac{1}{(2\pi)^{3/2} (2k_0)^{1/2}} \left[
\langle L, 0,S,\nu   | J,\nu \rangle \, \zeta_0 Z_0(k) + \frac{1}{2}
\langle L, 1,S,\nu-1 | J,\nu \rangle \, \zeta_+ Z_-(k) \right.
\nonumber\\
&& + \left. \frac{1}{2}
\langle L,-1,S,\nu+1 | J,\nu \rangle \, \zeta_- Z_+(k) \right] ~.
\label{anu}
\ea
Here $\alpha$ denotes the labels that,
in addition to $L$, $S$, $J$ and $\nu$, are needed to specify the
baryon wave function (see Eq.~(\ref{wf})).
The coefficients $\zeta_m$ are the spin-flavor matrix
elements of $X^M_{3} \, s_{3,m}$, to be discussed below,
and $Z_m(k)$ ($m=0,\pm$) are the radial matrix elements
\ba
Z_0(k) &=& 6 \int d \beta \, g(\beta) \,
\langle \alpha^{\prime},L^{\prime}=M_L^{\prime}=0 \, | \,
g k \, \hat U - h \, (\hat T_z - \frac{1}{6}(P_z+P^{\prime}_z) \hat U) \,
| \, \alpha,L,M_L=0 \rangle
\nonumber\\
&=& 6 \, [gk + h\frac{1}{6}(P_z+P_z^{\prime})] \, {\cal F}(k)
- 6h \, {\cal G}_z(k) ~,
\nonumber\\
Z_{\pm}(k) &=& -6h \int d \beta \, g(\beta) \,
\langle \alpha^{\prime},L^{\prime}=M_L^{\prime}=0 \, | \,
\hat T_{\pm} \, | \, \alpha,L,M_L=\mp 1 \rangle
\nonumber\\
&=& -6h \, {\cal G}_{\pm}(k) ~. \label{radme}
\ea
The calculation of the radial matrix elements is identical to that
already reported for electromagnetic couplings \cite{BIL,emff}.
Therefore, we do not repeat it here, but only show in Table~\ref{collff}
the matrix elements ${\cal F}(k)$ and ${\cal G}_m(k)$
of $\hat U$ and $\hat T_m$, respectively, for the `collective'
model with distribution given by Eq.~(\ref{gbeta}).
Note that Table~\ref{collff} presents the results
for an emission process, whereas the corresponding table for the
`collective' model in \cite{BIL,emff} shows the results for an
absorption process. For any other model of baryons with the same
spin-flavor structure, the corresponding results can be obtained
by replacing Table~\ref{collff} with the appropriate table
(for example, by using harmonic oscillator wave functions as
discussed in \cite{BIL}).

Contrary to the case of electromagnetic couplings where the spin-flavor
part is relatively simple, the calculation of the spin-flavor part for
strong decays is somewhat more involved. The spin-flavor matrix elements
factorize into a spin matrix element of $s_{3,m}$ between spin wave
functions and a flavor matrix element of $X_3^M$ between flavor wave
functions ($\phi$ and $\phi^{\prime}$). The calculation of the spin
part is straightforward. For the flavor part we take the flavor
operators of the form
\ba
X^{\pi^+} &=& - \frac{1}{\sqrt{2}} (\lambda_1 - i \, \lambda_2) ~,
\nonumber\\
X^{\pi^0} &=& \lambda_3 ~,
\nonumber\\
X^{\pi^-} &=& \frac{1}{\sqrt{2}} (\lambda_1 + i \, \lambda_2) ~,
\nonumber\\
X^{\eta_8} &=& \lambda_8 ~,
\nonumber\\
X^{\eta_1} &=& \sqrt{\frac{2}{3}} \, {\cal I} ~,
\label{meson}
\ea
where $\lambda_i$ are the Gell-Mann matrices \cite{PDG} and
${\cal I}$ denotes the unit operator in flavor space.
For the pseudoscalar $\eta$ mesons we introduce a mixing
angle $\theta_P$ between the octet and singlet mesons \cite{PDG}
\ba
\eta &=& \eta_8 \, \cos \theta_P - \eta_1 \, \sin \theta_P ~,
\nonumber\\
\eta^{\prime} &=& \eta_8 \, \sin \theta_P + \eta_1 \, \cos \theta_P ~,
\label{etamix}
\ea
and similarly for the corresponding flavor operators
\ba
X^{\eta} &=& X^{\eta_8} \, \cos \theta_P - X^{\eta_1} \, \sin \theta_P ~,
\nonumber\\
X^{\eta^{\prime}} &=& X^{\eta_8} \, \sin \theta_P
+ X^{\eta_1} \, \cos \theta_P ~.
\ea
The flavor operators $X^M_3$ appearing in Eq.~(\ref{hstrong})
only act on the 3-rd constituent.
The corresponding matrix elements can either be evaluated
explicitly for each channel separately, or more conveniently,
by using the Wigner-Eckart theorem and isoscalar factors of
$SU_f(3)$ \cite{Swart}.
For the decay process $B \rightarrow B^{\prime} + M$ we have
\ba
\langle \phi^{\prime} \, | \, X^M_{3} \,
| \, \phi \rangle &=&
\langle (p_2,q_2),I_2,M_{I_2},Y_2 \, | \, T^{(p,q),I,M_{I},Y} \,
| \, (p_1,q_1),I_1,M_{I_1},Y_1 \rangle
\nonumber\\
&=& \langle I_1,M_{I_1},I,M_{I}|I_2,M_{I_2} \rangle
\sum_{\gamma} \left< \left. \begin{array}{cc}
(p_1,q_1) & (p,q) \\ I_1,Y_1 & I,Y \end{array}
\right| \begin{array}{c} (p_2,q_2)_{\gamma} \\ I_2,Y_2 \end{array} \right>
\nonumber\\
&& \times \, \langle (p_2,q_2) \, || \, T^{(p,q)} \,
|| \, (p_1,q_1) \rangle_{\gamma} ~. \label{fme}
\ea
In this notation $(p_i,q_i)=(1,1)$ or $(3,0)$ for the baryon flavor octet
and decuplet, respectively, and $(p,q)=(1,1)$ or $(0,0)$ for the meson
flavor octet and singlet, respectively (see Eq.~(\ref{meson})).
The sum over $\gamma$ is over different multiplicities.
The flavor states are labeled by the quantum numbers $(p,q),I,M_I,Y$
corresponding to the reduction
$SU_f(3) \supset SU_I(2) \otimes U_Y(1)$.
The right hand side contains the sum of products of an isospin
Clebsch-Gordan coefficient, which contains the dependence on the
charge channel, a $SU(3)$ isoscalar factor, which
depends on the isospin channel, and a $SU(3)$ reduced matrix
element, which depends on the coupling of the flavor multiplets
$(p,q)$, but not on isospin $I,M_I$ and hypercharge $Y$.
In Table~\ref{su3rme} we give the expressions for
the $SU_f(3)$ reduced matrix elements.

Using Eqs.~(\ref{meson})--(\ref{fme}) and the matrix elements of
the spin operator, we can compute all spin-flavor matrix
elements for a given isospin channel.
In Tables~\ref{pi} and~\ref{eta} we present the results for
the decay into $\pi$ and $\eta$, $\eta^{\prime}$, respectively.
Results for a specific charge channel can be obtained by multiplying
with the appropriate isospin Clebsch-Gordan coefficient.

The helicity amplitudes $A_{\nu}(k)$ of Eq.~(\ref{anu}) can be
converted to partial wave amplitudes $a_l(k)$ by \cite{Rosner}
\ba
A_{\nu}(k) &=& \sum_{l=L \pm 1} \sqrt{2l+1} \,
\langle l,0,J^{\prime},\nu | J,\nu \rangle \, a_l(k) ~,
\nonumber\\
a_l(k) &=& \frac{1}{2J+1} \sum_{\nu} \sqrt{2l+1} \,
\langle l,0,J^{\prime},\nu | J,\nu \rangle \, A_{\nu}(k) ~.
\ea
Here $l$ is the relative orbital angular momentum between the final
baryon and the emitted meson. It takes the values $l=L \pm 1$ (the
value $l=L$ is not allowed because of parity conservation).
With the definition of the transition operator in Eq.~(\ref{hs}) and the
helicity amplitudes and partial wave amplitudes, the decay widths for a
specific isospin channel are given by \cite{LeYaouanc}
\ba
\Gamma(B \rightarrow B^{\prime} + M) &=& 2 \pi \rho_f \,
\frac{2}{2J+1} \sum_{\nu>0} | A_{\nu}(k) |^2
\nonumber\\
&=& 2 \pi \rho_f \, \sum_{l=L \pm 1} | a_l(k) |^2 ~,
\label{dw}
\ea
where $\rho_f$ is a phase space factor.

\section{Results}
\label{sec3}

For all resonances with the same value of $(v_1,v_2),L^P$
the expression for the decay widths of Eq.~(\ref{dw}) can be rewritten
in a more transparent form in terms of only two elementary partial
wave amplitudes $W_l(k)$,
\ba
\Gamma(B \rightarrow B^{\prime} + M)
\;=\; 2 \pi \rho_f \, \sum_{l=L \pm 1} | a_l(k) |^2
\;=\; 2 \pi \rho_f \, \frac{1}{(2\pi)^3 2k_0} \,
\sum_{l=L \pm 1} c_l \left| W_l(k) \right|^2 ~. \label{width}
\ea
For this set of resonances, the $k$ dependence of the partial wave
amplitudes $a_l(k)$ is contained in the amplitudes $W_l(k)$, while the
dependence on the individual baryon resonance is contained in the
coefficients $c_l$. In the algebraic method,
the elementary partial wave amplitudes $W_l(k)$ can be obtained in
closed form.

In Table~\ref{negwidth} we present the values of $c_l$ for the negative
parity resonances with $(v_1,v_2),L^P=(0,0),1^-$. In the `collective'
model with distribution given by Eq.~(\ref{gbeta}) the corresponding
$S$ and $D$ elementary partial wave amplitudes are
\ba
W_0(k) &=& i \, \left\{
[gk-\frac{1}{6}hk] \frac{ka}{(1+k^2a^2)^2}
+h \, m_3k_0a \frac{3-k^2a^2}{(1+k^2a^2)^3} \right\} ~,
\nonumber\\
W_2(k) &=& i \, \left\{
[gk-\frac{1}{6}hk] \frac{ka}{(1+k^2a^2)^2}
-h \, m_3k_0a \frac{4k^2a^2}{(1+k^2a^2)^3} \right\} ~. \label{pw02}
\ea
Similarly, in Table~\ref{poswidth} we present the $c_l$ coefficients
for the positive parity resonances with $(v_1,v_2),L^P=(0,0),2^+$.
The corresponding $P$ and $F$ elementary partial wave amplitudes are
\ba
W_1(k) &=& [gk-\frac{1}{6}hk]
\left\{ \frac{-1}{(1+k^2a^2)^2} + \frac{3}{2k^3a^3} H(ka) \right\}
\nonumber\\
&& +h \, m_3k_0a \frac{1}{ka} \frac{4k^2a^2}{(1+k^2a^2)^3} ~,
\nonumber\\
W_3(k) &=& [gk-\frac{1}{6}hk]
\left\{ \frac{-1}{(1+k^2a^2)^2} + \frac{3}{2k^3a^3} H(ka) \right\}
\nonumber\\
&&+h \, m_3k_0a \frac{1}{ka} \left\{ \frac{5+9k^2a^2}{(1+k^2a^2)^3}
- \frac{15}{2k^3a^3} H(ka) \right\} ~, \label{pw13}
\ea
with $H(ka) = \arctan ka - ka/(1+k^2a^2)$~.

Partial widths for other models of the nucleon and its resonances
can be obtained by introducing the corresponding expressions for the
elementary amplitudes $W_l(k)$. For example, the relevant
expressions in the harmonic oscillator quark model are
\ba
W_0(k) &=& \frac{i}{3} \left\{ [gk-\frac{1}{6}hk] k \beta
+ h m_3 k_0 \beta (3-\frac{k^2 \beta^2}{3}) \right\}
\mbox{e}^{-k^2 \beta^2/6} ~,
\nonumber\\
W_2(k) &=& \frac{i}{3} \left\{ [gk-\frac{1}{6}hk] k \beta
- \frac{1}{3} h m_3 k_0 \beta k^2 \beta^2 \right\}
\mbox{e}^{-k^2 \beta^2/6} ~,
\ea
and
\ba
W_1(k) &=& \frac{\sqrt{2}}{3\sqrt{15}} k \beta
\left\{ [gk-\frac{1}{6}hk] k \beta
+ h m_3 k_0 \beta (5-\frac{k^2 \beta^2}{3}) \right\}
\mbox{e}^{-k^2 \beta^2/6} ~,
\nonumber\\
W_3(k) &=& \frac{\sqrt{2}}{3\sqrt{15}} k \beta
\left\{ [gk-\frac{1}{6}hk] k \beta
- \frac{1}{3} h m_3 k_0 \beta k^2 \beta^2 \right\}
\mbox{e}^{-k^2 \beta^2/6} ~.
\ea

\section{Analysis of experimental data}
\label{sec4}

Use of Eqs.~(\ref{width})--(\ref{pw13}) allows us to do a straightforward
and systematic analysis of the experimental data.
Here we adopt the procedure of \cite{LeYaouanc}, in which
calculations are performed in the rest frame of the decaying resonance,
and in which the relativistic expression for the phase space factor
$\rho_f$ as well as for the momentum $k$ of the emitted meson are retained.
The expressions for $k$ and $\rho_f$ are
\ba
k^2 &=& -m_M^2 + \frac{(m_B^2-m_{B^{\prime}}^2+m_M^2)^2}{4m_B^2} ~,
\nonumber\\
\rho_f &=& \int d \vec{P}^{\prime} \, d \vec{P}_M \,
\delta(m_B-E_{B^{\prime}}(P_M)-E_M(P_M))
\, \delta(\vec{P}^{\prime}+\vec{P}_M)
\nonumber\\
&=& 4 \pi \frac{E_{B^{\prime}}(k) E_M(k) k}{m_B}
\ea
with $E_{B^{\prime}}(k)=\sqrt{m_{B^{\prime}}^2+k^2}$ and
$E_{M}(k)=\sqrt{m_{M}^2+k^2}$.

We consider here decays with emission of $\pi$ and $\eta$.
The experimental data, extracted from the
compilation by the Particle Data Group \cite{PDG} are shown in
Tables~\ref{npi},~\ref{dpi} and~\ref{ndeta}, where they are compared
with the results of our calculation. The calculated values depend on the
two parameters $g$ and $h$ in the transition operator of Eq.~(\ref{hs}),
and on the scale parameter $a$ of Eq.~(\ref{gbeta}).
We keep $g$, $h$ and $a$ equal
for {\em all} resonances and {\em all} decay channels ($N \pi$,
$N \eta$, $\Delta \pi$, $\Delta \eta$). In comparing with previous
calculations, it should be noted that in the
calculation in the nonrelativistic quark model of \cite{KI}
the decay widths are parametrized by four reduced partial wave amplitudes
instead of the two elementary amplitudes $g$ and $h$. Furthermore,
the momentum dependence of these reduced amplitudes are represented
by constants. The calculation in the relativized quark model of \cite{CR}
was done using a pair-creation model for the decay and
involved a different assumption on the phase space factor.
Both the nonrelativistic and relativized quark model calculations
include the effects of mixing induced by the hyperfine interaction,
which in the present calculation are not taken into account.

In the present analysis we determine the
values of $g$, $h$ and $a$ from a least square fit to the $N \pi$
partial widths (which are relatively well known) with the exclusion
of the $S_{11}$ resonances. For the latter the situation is not clear due
to possible mixing of $N(1535)S_{11}$ and $N(1650)S_{11}$ and
the possible existence of a third $S_{11}$ resonance \cite{LW}. As a
result we find $g=1.164$ GeV$^{-1}$ and $h=-0.094$ GeV$^{-1}$. The
relative sign is consistent with a previous analysis of the strong
decay of mesons \cite{GIK} and with a derivation from the axial-vector
coupling (see {\it e.g.} \cite{LeYaouanc}).
The scale parameter, $a=0.232$ fm, extracted in the present fit
is found to be equal to the value extracted in the calculation of
electromagnetic couplings \cite{emff}.
Just as in our study of the electromagnetic couplings \cite{emff}
we present the strong decays of the resonances, that in the `collective'
model are assigned as vibrational excitations of the configuration
of Fig.~\ref{geometry}, in terms of two coefficients,
$\chi_1=(1-R^2)/R\sqrt{N}$ and $\chi_2=\sqrt{1+R^2}/R\sqrt{N}$,
one for each fundamental vibration (here $N$ determines the size
of the model space and $R^2$ is a size parameter, as discussed in
\cite{BIL}).

The calculation of decay widths into the $N \pi$ channel,
as shown for the 3 and 4 star resonances in Table~\ref{npi}, is in
fair agreement with experiment. This is emphasized in
Figs.~\ref{nwidths} and~\ref{dwidths}. The results are to a large
extent a consequence of spin-flavor symmetry. The use of `collective'
form factors improves somewhat the results when compared with older
(harmonic oscillator) calculations. This is shown in Table~\ref{regge}
where the decay of a $\Delta$ Regge trajectory into $N \pi$
is analyzed and compared with the calculations of \cite{LeYaouanc},
which are based on the harmonic oscillator model discussed in \cite{Dalitz}.
We also include the results of more recent calculations in the
nonrelativistic quark model \cite{KI} and in the relativized quark model
\cite{CR}. There does not seem to be anything unusual in the decays into
$\pi$ and our analysis confirms the results of previous analyses.
In Table~\ref{ndmiss} we show the $N \pi$ decays of the
calculated resonances below 2 GeV which have not
been observed (missing resonances). The resonances with $L^P=1^+$
and $L^P=2^-$ are decoupled because of the spin-flavor symmetry.
Most of other resonances have small decay widths, with the exception
of the $P_{13}$ and $F_{15}$ states, which have widths comparable
to those of the well-established $P_{13}$ and $F_{15}$ states of
Table~\ref{npi}. This behaviour can be understood
by inspection of Table~\ref{poswidth}. The coefficients for the
$P_{13}$ states are $c_1=25/12$ and $c_1=2$.
This leads to comparable $N \pi$ widths of 31 MeV and 56 MeV,
respectively. The difference is due to the $k$ values of these resonances
that enter in the elementary partial wave amplitude $W_1(k)$ of
Eq.~(\ref{pw13}). A similar situation holds for the $F_{15}$ states.
The relative magnitude of the $c_l$ coefficients in Table~\ref{poswidth}
explain the smaller decay widths for the other missing resonances of
Table~\ref{ndmiss}.

Contrary to the decays into $\pi$,
the decay widths into $\eta$ have some unusual
properties. The calculation gives systematically small values
for these widths. This is due to a combination of phase space
factors and the structure of the transition operator. Both depend
on the momentum transfer $k$.
The values of $k$ in Tables~\ref{npi},~\ref{dpi}
and~\ref{ndeta} show that, due to the difference between
the $\pi$ and $\eta$ mass, the momentum carried by the $\eta$ is
smaller than that carried by the $\pi$. Therefore, the $\eta$ decay widths
are suppressed relative to the $\pi$ decays. The spin-flavor part is
approximately the same for $N \pi$ and $N \eta$, since $\pi$ and $\eta$
are in the same $SU_f(3)$ multiplet. We emphasize here, that
the transition operator was determined
by fitting the coefficients $g$ and $h$ to the $N \pi$ decays
of the 3 and 4 star resonances. Hence the $\eta$ decays are
calculated without introducing any further parameters.

The experimental situation is unclear. The 1992 PDG compilation
\cite{PDG92} gave systematically small widths ($\sim 1$ MeV) for all
resonances except $N(1535)S_{11}$. The 1994 PDG compilation \cite{PDG94}
deleted all $\eta$ widths with the exception of $N(1535)S_{11}$.
This situation persists in the latest PDG compilation \cite{PDG},
where $N(1650)S_{11}$ is now assigned a small but non-zero $\eta$ width.
The results of our analysis suggest that the large $\eta$ width for the
$N(1535)S_{11}$ is not due to a conventional $q^3$ state. One possible
explanation is the presence of another state in the same mass region,
{\it e.g.} a quasi-bound meson-baryon $S$ wave resonance just below
or above threshold, for example $N\eta$, $K\Sigma$ or $K\Lambda$
\cite{Kaiser}. Another possibility is an exotic configuration of four
quarks and one antiquark ($q^{4}\bar{q}$).

For possible use in the analysis of new experimental data,
we give in Table~\ref{ndmiss} the strong decay widths of the socalled
missing resonances with a calculated mass below 2 GeV
which up to now have not been observed experimentally.

\section{Conclusions}
\label{sec5}

We have presented a calculation of the strong decay widths
$N^{\ast} \rightarrow N + \pi$,
$N^{\ast} \rightarrow \Delta + \pi$,
$N^{\ast} \rightarrow N + \eta$,
$\Delta^{\ast} \rightarrow N + \pi$,
$\Delta^{\ast} \rightarrow \Delta + \pi$ and
$\Delta^{\ast} \rightarrow \Delta + \eta$ in a collective model of baryons.
By exploiting the symmetry of the problem, both in its spin-flavor-color
part, $SU_{sf}(6) \otimes SU_c(3)$, and in its space part, $U(7)$, we
have been able to write the results in a transparent analytic way
(Section~\ref{sec3}). The analysis of experimental data shows that,
while the decays into $\pi$ follow the expected pattern, the decays
into $\eta$ have some unusual features.
Our calculations do not show any indication for a large $\eta$ width,
as is observed for the $N(1535)S_{11}$ resonance. The observed large
$\eta$ width indicates the presence of another configuration, which is
outside the present model space. This suggests, that
in order to elucidate this point, particular attention be paid at
CEBAF to the $N \eta$ channel.

Our calculations can be easily
extended to include other decay channels, such as $\Lambda K$ and
$\Sigma K$. These calculations are currently underway and are part
of the extension of the model to include strange resonances as well.

\section*{Acknowledgements}

We wish to thank T.-S. H. Lee and D. Kurath for interesting discussions.
This work is supported in part
by CONACyT, M\'exico under project 400340-5-3401E, DGAPA-UNAM under
project IN105194 (R.B.), by D.O.E. Grant DE-FG02-91ER40608 (F.I.) and
by grant No. 94-00059 from the United States-Israel Binational Science
Foundation (BSF), Jerusalem, Israel (A.L.).

\clearpage
\begin{table}
\centering
\caption[]{\small
Matrix elements ${\cal F}(k)$ and ${\cal G}(k)$ in the collective
model for $N \rightarrow \infty$ (large model space).
The final state is $[56,0^+]_{(0,0);0}$. The matrix elements for the
vibrational excitations are given in terms of $\chi_1=(1-R^2)/R\sqrt{N}$
and $\chi_2=\sqrt{1+R^2}/R\sqrt{N}$ \cite{BIL,emff}.
$H(x)=\arctan x - x/(1+x^2)$.
\normalsize}
\label{collff} \vspace{15pt}
\begin{tabular}{cccc}
& & & \\
Initial state & ${\cal F}(k)$
& ${\cal G}_{z}(k)/m_3 k_0 a$ & ${\cal G}_{\pm}(k)/m_3 k_0 a$ \\
& & & \\
\hline
& & & \\
$[56,0^+]_{(0,0);0}$
& $\frac{1}{(1+k^2a^2)^2}$ & $\frac{4ka}{(1+k^2a^2)^3}$ & 0  \\
& & & \\
$[20,1^+]_{(0,0);0}$ & 0 & 0 & 0 \\
& & & \\
$[70,1^-]_{(0,0);1}$ & $i \, \sqrt{3} \, \frac{ka}{(1+k^2a^2)^2}$
& $-i \, \sqrt{3} \, \frac{1-3k^2a^2}{(1+k^2a^2)^3}$
& $\mp i \, \sqrt{6} \, \frac{1}{(1+k^2a^2)^2}$ \\
& & & \\
$[56,2^+]_{(0,0);0}$
& $\frac{1}{2}\sqrt{5}\left[ \frac{-1}{(1+k^2a^2)^2} \right.$
& $-\frac{1}{2}\sqrt{5}\left[ \frac{3+7k^2a^2}{ka(1+k^2a^2)^3} \right.$
& $\mp \sqrt{\frac{15}{2}}\left[ \frac{-1}{ka(1+k^2a^2)^2} \right.$ \\
& $\left. \hspace{1cm} + \frac{3}{2k^3a^3} H(ka) \right]$
& $\left. \hspace{1cm} - \frac{9}{2k^4a^4} H(ka) \right]$
& $\left. \hspace{1cm} + \frac{3}{2k^4a^4} H(ka) \right]$ \\
& & & \\
$[70,2^-]_{(0,0);1}$ & 0 & 0 & 0 \\
& & & \\
$[70,2^+]_{(0,0);2}$
& $-\frac{1}{2}\sqrt{15}\left[ \frac{-1}{(1+k^2a^2)^2} \right.$
& $\frac{1}{2}\sqrt{15}\left[ \frac{3+7k^2a^2}{ka(1+k^2a^2)^3} \right.$
& $\pm \frac{3}{2}\sqrt{10}\left[ \frac{-1}{ka(1+k^2a^2)^2} \right.$ \\
& $\left. \hspace{1cm} + \frac{3}{2k^3a^3} H(ka) \right]$
& $\left. \hspace{1cm} - \frac{9}{2k^4a^4} H(ka) \right]$
& $\left. \hspace{1cm} + \frac{3}{2k^4a^4} H(ka) \right]$ \\
& & & \\
\hline
& & & \\
$[56,0^+]_{(1,0);0}$
& $-\chi_1 \frac{2k^2a^2}{(1+k^2a^2)^3}$
& $ \chi_1 \frac{4ka(1-2k^2a^2)}{(1+k^2a^2)^4}$ & 0 \\
& & & \\
$[70,1^-]_{(1,0);1}$
& $ i \, \frac{1}{2}\sqrt{3} \, \chi_1 \frac{ka(1-3k^2a^2)}{(1+k^2a^2)^3}$
& $-i \, \frac{1}{2}\sqrt{3} \,
\chi_1 \frac{1-14k^2a^2+9k^4a^4}{(1+k^2a^2)^4}$
& $\mp i \, \sqrt{\frac{3}{2}} \, \chi_1 \frac{1-3k^2a^2}{(1+k^2a^2)^3}$ \\
& & & \\
\hline
& & & \\
$[70,0^+]_{(0,1);0}$
& $ \chi_2 \frac{2k^2a^2}{(1+k^2a^2)^3}$
& $-\chi_2 \frac{4ka(1-2k^2a^2)}{(1+k^2a^2)^4}$ & 0 \\
& & & \\
$[70,1^-]_{(0,1);1}$
& $-i \, \sqrt{\frac{3}{2}} \, \chi_2 \frac{ka(1-k^2a^2)}{(1+k^2a^2)^3}$
& $ i \, \sqrt{\frac{3}{2}} \, \chi_2 \frac{1-8k^2a^2+3k^4a^4}{(1+k^2a^2)^4}$
& $\pm i \, \sqrt{3} \, \chi_2 \frac{1-k^2a^2}{(1+k^2a^2)^3}$ \\
& & & \\
$[56,1^-]_{(0,1);1}$
& $-i \, \sqrt{6} \, \chi_2 \frac{k^3a^3}{(1+k^2a^2)^3}$
& $ i \, \sqrt{6} \, \chi_2 \frac{3k^2a^2(1-k^2a^2)}{(1+k^2a^2)^4}$
& $\pm i \, 2\sqrt{3} \, \chi_2 \frac{k^2a^2}{(1+k^2a^2)^3}$ \\
& & & \\
\end{tabular}
\end{table}

\clearpage
\begin{table}
\centering
\caption[]{\small
$SU_f(3)$ reduced matrix elements.
\normalsize}
\label{su3rme}
\vspace{15pt}
\begin{tabular}{cccc}
& & & \\
\multicolumn{4}{c}
{$< (p_2,q_2) \, || \, T^{(1,1)} \, || \, (p_1,q_1) >_{\gamma}$} \\
$(p_2,q_2)$ & $(p_1,q_1)$ & $\gamma=1$ & $\gamma=2$ \\
& & & \\
\hline
& & & \\
$(1,1)_{\lambda}$ & $(1,1)_{\lambda}$
& $-\frac{\sqrt{5}}{\sqrt{3}}$ & $\frac{1}{\sqrt{3}}$ \\
& & & \\
$(1,1)_{\rho}$ & $(1,1)_{\rho}$
& $ \frac{\sqrt{5}}{\sqrt{3}}$ & $\sqrt{3}$ \\
& & & \\
$(3,0)$ & $(3,0)$ & $\frac{2\sqrt{2}}{\sqrt{3}}$ & \\
& & & \\
$(3,0)$ & $(1,1)_{\lambda}$ & $\frac{2\sqrt{2}}{\sqrt{3}}$ & \\
& & & \\
$(3,0)$ & $(1,1)_{\rho}$ & $0$ & \\
& & & \\
$(1,1)_{\lambda}$ & $(3,0)$ & $-\frac{\sqrt{10}}{\sqrt{3}}$ & \\
& & & \\
$(1,1)_{\rho}$ & $(3,0)$ & $0$ & \\
& & & \\
\end{tabular}
\end{table}

\clearpage
\begin{table}
\centering
\caption[]{\small
Spin-flavor matrix elements $\zeta_m$ ($m=0,\pm$) of Eq.~(\ref{anu}) for
strong decay of baryons $B \rightarrow B^{\prime} + M$ with $M=\pi$.
The final state is $^{2}8_{1/2}[56,0^+]$ for the nucleon ($B^{\prime}=N$)
and $^{4}10_{3/2}[56,0^+]$ for the delta ($B^{\prime}=\Delta$).
\normalsize}
\label{pi}
\vspace{15pt}
\begin{tabular}{c|ccc|ccc|ccc}
& & & & & & & & & \\
& \multicolumn{3}{c|} {$A_{1/2}(N \pi)$}
& \multicolumn{3}{c|} {$A_{1/2}(\Delta \pi)$}
& \multicolumn{3}{c}  {$A_{3/2}(\Delta \pi)$} \\
Initial State & $\zeta_0$ & $\zeta_+$ & $\zeta_-$
& $\zeta_0$ & $\zeta_+$ & $\zeta_-$
& $\zeta_0$ & $\zeta_+$ & $\zeta_-$ \\
& & & & & & & & & \\
\hline
& & & & & & & & & \\
$^{2}8_J[56,L^P]$ & $\frac{5}{6\sqrt{3}}$ & $\frac{5}{3\sqrt{3}}$ & 0
& $\frac{-2\sqrt{2}}{3\sqrt{3}}$ & $\frac{2\sqrt{2}}{3\sqrt{3}}$ & 0
& 0 & $\frac{2\sqrt{2}}{3}$ & 0 \\
& & & & & & & & & \\
$^{2}8_J[70,L^P]$ & $\frac{\sqrt{2}}{3\sqrt{3}}$
& $\frac{2\sqrt{2}}{3\sqrt{3}}$ & 0
& $\frac{2}{3\sqrt{3}}$ & $\frac{-2}{3\sqrt{3}}$ & 0
& 0 & $\frac{-2}{3}$ & 0 \\
& & & & & & & & & \\
$^{4}8_J[70,L^P]$ & $\frac{1}{3\sqrt{6}}$ & $\frac{1}{3\sqrt{6}}$
& $\frac{-1}{3\sqrt{2}}$
& $\frac{1}{3\sqrt{3}}$ & $\frac{4}{3\sqrt{3}}$ & $\frac{2}{3}$
& $\frac{1}{\sqrt{3}}$ & $\frac{2}{3}$ & 0 \\
& & & & & & & & & \\
$^{2}8_J[20,L^P]$ & 0 & 0 & 0 & 0 & 0 & 0 & 0 & 0 & 0 \\
& & & & & & & & & \\
\hline
& & & & & & & & & \\
$^{4}10_J[56,L^P]$ & $\frac{2}{3\sqrt{3}}$ & $\frac{2}{3\sqrt{3}}$
& $\frac{-2}{3}$
& $\frac{\sqrt{5}}{6\sqrt{3}}$ & $\frac{2\sqrt{5}}{3\sqrt{3}}$
& $\frac{\sqrt{5}}{3}$
& $\frac{\sqrt{5}}{2\sqrt{3}}$ & $\frac{\sqrt{5}}{3}$ & 0 \\
& & & & & & & & & \\
$^{2}10_J[70,L^P]$ & $\frac{1}{6\sqrt{3}}$ & $\frac{1}{3\sqrt{3}}$ & 0
& $\frac{-\sqrt{5}}{3\sqrt{3}}$ & $\frac{\sqrt{5}}{3\sqrt{3}}$ & 0
& 0 & $\frac{\sqrt{5}}{3}$ & 0 \\
& & & & & & & & & \\
\end{tabular}
\end{table}

\clearpage
\begin{table}
\centering
\caption[]{\small
Same as Table~\ref{pi}, but with $M=\eta$ and $M=\eta^{\prime}$ corresponding
to\hfil\break
$\xi=(\cos \theta_P - \sqrt{2} \sin \theta_P)/\sqrt{3}$ and
$\xi=(\sin \theta_P + \sqrt{2} \cos \theta_P)/\sqrt{3}$, respectively.
\normalsize}
\label{eta}
\vspace{15pt}
\begin{tabular}{c|ccc|ccc|ccc}
& & & & & & & & & \\
& \multicolumn{3}{c|} {$A_{1/2}(N \eta)/\xi$}
& \multicolumn{3}{c|} {$A_{1/2}(\Delta \eta)/\xi$}
& \multicolumn{3}{c}  {$A_{3/2}(\Delta \eta)/\xi$} \\
Initial State & $\zeta_0$ & $\zeta_+$ & $\zeta_-$
& $\zeta_0$ & $\zeta_+$ & $\zeta_-$
& $\zeta_0$ & $\zeta_+$ & $\zeta_-$ \\
& & & & & & & & & \\
\hline
& & & & & & & & & \\
$^{2}8_J[56,L^P]$ & $\frac{1}{6}$ & $\frac{1}{3}$ & 0
& 0 & 0 & 0 & 0 & 0 & 0 \\
& & & & & & & & & \\
$^{2}8_J[70,L^P]$ & $\frac{1}{3\sqrt{2}}$ & $\frac{\sqrt{2}}{3}$ & 0
& 0 & 0 & 0 & 0 & 0 & 0 \\
& & & & & & & & & \\
$^{4}8_J[70,L^P]$ & $\frac{-1}{3\sqrt{2}}$ & $\frac{-1}{3\sqrt{2}}$
& $\frac{1}{\sqrt{6}}$ & 0 & 0 & 0 & 0 & 0 & 0 \\
& & & & & & & & & \\
$^{2}8_J[20,L^P]$ & 0 & 0 & 0 & 0 & 0 & 0 & 0 & 0 & 0 \\
& & & & & & & & & \\
\hline
& & & & & & & & & \\
$^{4}10_J[56,L^P]$ & 0 & 0 & 0
& $\frac{1}{6}$ & $\frac{2}{3}$ & $\frac{1}{\sqrt{3}}$
& $\frac{1}{2}$ & $\frac{1}{\sqrt{3}}$ & 0 \\
& & & & & & & & & \\
$^{2}10_J[70,L^P]$ & 0 & 0 & 0
& $\frac{-1}{3}$ & $\frac{1}{3}$ & 0
& 0 & $\frac{1}{\sqrt{3}}$ & 0 \\
& & & & & & & & & \\
\end{tabular}
\end{table}

\clearpage
\begin{table}
\centering
\caption[]{\small
Coefficients $c_l$ of Eq.~(\ref{width}) for the strong decay widths
$B \rightarrow B^{\prime} + M$
of the negative parity resonances with $(v_1,v_2),L^P=(0,0),1^-$.
The final state is $^{2}8_{1/2}[56,0^+]_{(0,0);0}$ for the nucleon
($B^{\prime}=N$) and $^{4}10_{3/2}[56,0^+]_{(0,0);0}$ for the delta
($B^{\prime}=\Delta$). $\xi$ is defined in the caption of
Table~\ref{eta}.
\normalsize}
\label{negwidth}
\vspace{15pt}
\begin{tabular}{cc|cc|cc|cc|cc}
& & & & & & & & & \\
\multicolumn{2}{c|} {State} & \multicolumn{2}{c|} {$N \pi$}
& \multicolumn{2}{c|} {$N \eta$}
& \multicolumn{2}{c|} {$\Delta \pi$}
& \multicolumn{2}{c} {$\Delta \eta$} \\
& & $c_0$ & $c_2$ & $c_0$ & $c_2$ & $c_0$ & $c_2$ & $c_0$ & $c_2$ \\
& & & & & & & & & \\
\hline
& & & & & & & & & \\
$S_{11}$ & $^{2}8_{1/2}[70,1^-]_{(0,0);1}$
& $\frac{8}{3}$ & & $2 \xi^2$ & & & $\frac{16}{3}$ & & \\
& & & & & & & & & \\
$D_{13}$ & $^{2}8_{3/2}[70,1^-]_{(0,0);1}$
& & $\frac{8}{3}$ & & $2 \xi^2$ & $\frac{8}{3}$ & $\frac{8}{3}$ & & \\
& & & & & & & & & \\
$S_{11}$ & $^{4}8_{1/2}[70,1^-]_{(0,0);1}$
& $\frac{2}{3}$ & & $2 \xi^2$ & & & $\frac{4}{3}$ & & \\
& & & & & & & & & \\
$D_{13}$ & $^{4}8_{3/2}[70,1^-]_{(0,0);1}$
& & $\frac{1}{15}$ & & $\frac{1}{5} \xi^2$ & $\frac{20}{3}$
& $\frac{64}{15}$ & & \\
& & & & & & & & & \\
$D_{15}$ & $^{4}8_{5/2}[70,1^-]_{(0,0);1}$
& & $\frac{2}{5}$ & & $\frac{6}{5} \xi^2$ & & $\frac{28}{5}$ & & \\
& & & & & & & & & \\
\hline
& & & & & & & & & \\
$S_{31}$ & $^{2}10_{1/2}[70,1^-]_{(0,0);1}$
& $\frac{1}{3}$ & & & & & $\frac{20}{3}$ & & $4 \xi^2$ \\
& & & & & & & & & \\
$D_{33}$ & $^{2}10_{3/2}[70,1^-]_{(0,0);1}$
& & $\frac{1}{3}$ & & & $\frac{10}{3}$ & $\frac{10}{3}$
& $2 \xi^2$ & $2 \xi^2$ \\
& & & & & & & & & \\
\end{tabular}
\end{table}

\clearpage
\begin{table}
\centering
\caption[]{\small
Coefficients $c_l$ of Eq.~(\ref{width}) for the strong decay widths
of the positive parity resonances with $(v_1,v_2),L^P=(0,0),2^+$.
For notation see Table~\ref{negwidth}.
\normalsize}
\label{poswidth}
\vspace{15pt}
\begin{tabular}{cc|cc|cc|cc|cc}
& & & & & & & & & \\
\multicolumn{2}{c|} {State} & \multicolumn{2}{c|} {$N \pi$}
& \multicolumn{2}{c|} {$N \eta$}
& \multicolumn{2}{c|} {$\Delta \pi$}
& \multicolumn{2}{c} {$\Delta \eta$} \\
& & $c_1$ & $c_3$ & $c_1$ & $c_3$ & $c_1$ & $c_3$ & $c_1$ & $c_3$ \\
& & & & & & & & & \\
\hline
& & & & & & & & & \\
$P_{13}$ & $^{2}8_{3/2}[56,2^+]_{(0,0);0}$
& $\frac{25}{12}$ & & $\frac{1}{4} \xi^2$ & & $\frac{4}{15}$
& $\frac{12}{5}$ & & \\
& & & & & & & & & \\
$F_{15}$ & $^{2}8_{5/2}[56,2^+]_{(0,0);0}$
& & $\frac{25}{12}$ & & $\frac{1}{4} \xi^2$ & $\frac{8}{5}$
& $\frac{16}{15}$ & & \\
& & & & & & & & & \\
$P_{13}$ & $^{2}8_{3/2}[70,2^+]_{(0,0);2}$
& $2$ & & $\frac{3}{2} \xi^2$ & & $\frac{2}{5}$
& $\frac{18}{5}$ & & \\
& & & & & & & & & \\
$F_{15}$ & $^{2}8_{5/2}[70,2^+]_{(0,0);2}$
& & $2$ & & $\frac{3}{2} \xi^2$ & $\frac{12}{5}$
& $\frac{8}{5}$ & & \\
& & & & & & & & & \\
$P_{11}$ & $^{4}8_{1/2}[70,2^+]_{(0,0);2}$
& $\frac{1}{2}$ & & $\frac{3}{2} \xi^2$ & & $1$ & & & \\
& & & & & & & & & \\
$P_{13}$ & $^{4}8_{3/2}[70,2^+]_{(0,0);2}$
& $\frac{1}{4}$ & & $\frac{3}{4} \xi^2$ & & $\frac{16}{5}$
& $\frac{9}{5}$ & & \\
& & & & & & & & & \\
$F_{15}$ & $^{4}8_{5/2}[70,2^+]_{(0,0);2}$
& & $\frac{1}{14}$ & & $\frac{3}{14} \xi^2$ & $\frac{21}{5}$
& $\frac{128}{35}$ & & \\
& & & & & & & & & \\
$F_{17}$ & $^{4}8_{7/2}[70,2^+]_{(0,0);2}$
& & $\frac{9}{28}$ & & $\frac{27}{28} \xi^2$ &
& $\frac{27}{7}$ & & \\
& & & & & & & & & \\
\hline
& & & & & & & & & \\
$P_{31}$ & $^{4}10_{1/2}[56,2^+]_{(0,0);0}$
& $\frac{4}{3}$ & & & & $\frac{5}{12}$ & & $\frac{1}{4} \xi^2$ & \\
& & & & & & & & & \\
$P_{33}$ & $^{4}10_{3/2}[56,2^+]_{(0,0);0}$
& $\frac{2}{3}$ & & & & $\frac{4}{3}$ & $\frac{3}{4}$
& $\frac{4}{5} \xi^2$ & $\frac{9}{20} \xi^2$ \\
& & & & & & & & & \\
$F_{35}$ & $^{4}10_{5/2}[56,2^+]_{(0,0);0}$
& & $\frac{4}{21}$ & & & $\frac{7}{4}$ & $\frac{32}{21}$
& $\frac{21}{20} \xi^2$ & $\frac{32}{35} \xi^2$ \\
& & & & & & & & & \\
$F_{37}$ & $^{4}10_{7/2}[56,2^+]_{(0,0);0}$
& & $\frac{6}{7}$ & & & & $\frac{45}{28}$ & & $\frac{27}{28} \xi^2$ \\
& & & & & & & & & \\
$P_{33}$ & $^{2}10_{3/2}[70,2^+]_{(0,0);2}$
& $\frac{1}{4}$ & & & & $\frac{1}{2}$ & $\frac{9}{2}$
& $\frac{3}{10} \xi^2$ & $\frac{27}{10} \xi^2$ \\
& & & & & & & & & \\
$F_{35}$ & $^{2}10_{5/2}[70,2^+]_{(0,0);2}$
& & $\frac{1}{4}$ & & & $3$ & $2$
& $\frac{9}{5} \xi^2$ & $\frac{6}{5} \xi^2$ \\
& & & & & & & & & \\
\end{tabular}
\end{table}

\clearpage
\begin{table}
\centering
\caption[]{\small
$N \pi$ decay widths of (3 and 4 star) nucleon and delta
resonances in MeV. The experimental values are taken from \cite{PDG}.
The mixing angle for the $\eta$ mesons is $\theta_P=-23^{\circ}$
\cite{GIK}. $\chi_1$ and $\chi_2$ are defined in the caption of
Table~\ref{collff}.
\normalsize}
\label{npi}
\vspace{15pt}
\begin{tabular}{lccccc}
& & & & & \\
State & Mass & Resonance & $k$(MeV) & $\Gamma$(th) & $\Gamma$(exp) \\
& & & & & \\
\hline
& & & & & \\
$S_{11}$ & $N(1535)$ & $^{2}8_{1/2}[70,1^-]_{(0,0);1}$
& $467$ &  $85$ & $79 \pm 38$ \\
$S_{11}$ & $N(1650)$ & $^{4}8_{1/2}[70,1^-]_{(0,0);1}$
& $547$ &  $35$ & $130 \pm 27$ \\
$P_{13}$ & $N(1720)$ & $^{2}8_{3/2}[56,2^+]_{(0,0);0}$
& $594$ &  $31$ & $22 \pm 11$ \\
$D_{13}$ & $N(1520)$ & $^{2}8_{3/2}[70,1^-]_{(0,0);1}$
& $456$ & $115$ & $67 \pm 9$ \\
$D_{13}$ & $N(1700)$ & $^{4}8_{3/2}[70,1^-]_{(0,0);1}$
& $580$ &   $5$ & $10 \pm 7$ \\
$D_{15}$ & $N(1675)$ & $^{4}8_{5/2}[70,1^-]_{(0,0);1}$
& $564$ &  $31$ & $72 \pm 12$ \\
$F_{15}$ & $N(1680)$ & $^{2}8_{5/2}[56,2^+]_{(0,0);0}$
& $567$ &  $41$ & $84 \pm 9$ \\
$G_{17}$ & $N(2190)$ & $^{2}8_{7/2}[70,3^-]_{(0,0);1}$
& $888$ &  $34$ & $67 \pm 27$ \\
$G_{19}$ & $N(2250)$ & $^{4}8_{9/2}[70,3^-]_{(0,0);1}$
& $923$ &  $7$ & $38 \pm 21$ \\
$H_{19}$ & $N(2220)$ & $^{2}8_{9/2}[56,4^+]_{(0,0);0}$
& $905$ &  $15$ & $65 \pm 28$ \\
$I_{1,11}$ & $N(2600)$ & $^{2}8_{11/2}[70,5^-]_{(0,0);1}$
& $1126$ &  $9$ & $49 \pm 20$ \\
& & & & & \\
$P_{11}$ & $N(1440)$ & $^{2}8_{1/2}[56,0^+]_{(1,0);0}$
& $398$ &  $108 \, \chi_1^2$ & $227 \pm 67$ \\
$P_{11}$ & $N(1710)$ & $^{2}8_{1/2}[70,0^+]_{(0,1);0}$
& $587$ &  $173 \, \chi_2^2$ & $22 \pm 17$ \\
& & & & & \\
\hline
& & & & & \\
$S_{31}$ & $\Delta(1620)$ & $^{2}10_{1/2}[70,1^-]_{(0,0);1}$
& $526$ &  $16$ & $37 \pm 11$ \\
$P_{31}$ & $\Delta(1910)$ & $^{4}10_{1/2}[56,2^+]_{(0,0);0}$
& $716$ &  $42$ & $52 \pm 19$ \\
$P_{33}$ & $\Delta(1232)$ & $^{4}10_{3/2}[56,0^+]_{(0,0);0}$
& $229$ & $116$ & $119 \pm 5$ \\
$P_{33}$ & $\Delta(1920)$ & $^{4}10_{3/2}[56,2^+]_{(0,0);0}$
& $723$ &  $22$ & $28 \pm 19$ \\
$D_{33}$ & $\Delta(1700)$ & $^{2}10_{3/2}[70,1^-]_{(0,0);1}$
& $580$ &  $27$ & $45 \pm 21$ \\
$D_{35}$ & $\Delta(1930)$ & $^{2}10_{5/2}[70,2^-]_{(0,0);1}$
& $729$ &   $0$ & $52 \pm 23$ \\
$F_{35}$ & $\Delta(1905)$ & $^{4}10_{5/2}[56,2^+]_{(0,0);0}$
& $713$ &   $9$ & $36 \pm 20$ \\
$F_{37}$ & $\Delta(1950)$ & $^{4}10_{7/2}[56,2^+]_{(0,0);2}$
& $741$ &  $45$ & $120 \pm 14$ \\
$H_{3,11}$ & $\Delta(2420)$ & $^{4}10_{11/2}[56,4^+]_{(0,0);0}$
& $1023$ &  $12$ & $40 \pm 22$ \\
& & & & & \\
$S_{31}$ & $\Delta(1900)$ & $^{2}10_{1/2}[70,1^-]_{(1,0);1}$
& $710$ &   $2 \, \chi_1^2$ & $38 \pm 21$ \\
$P_{33}$ & $\Delta(1600)$ & $^{4}10_{3/2}[56,0^+]_{(1,0);0}$
& $513$ &   $108 \, \chi_1^2$ & $61 \pm 32$ \\
& & & & & \\
\end{tabular}
\end{table}

\clearpage
\begin{table}
\centering
\caption[]{\small
$\Delta \pi$ decay widths of (3 and 4 star) nucleon and delta
resonances in MeV. For classification of the resonances see
Table~\ref{npi}. The experimental values are taken from \cite{PDG}.
\normalsize}
\label{dpi}
\vspace{15pt}
\begin{tabular}{lcccccccc}
& & & & & & & & \\
State & Mass & $k$(MeV) & $l$ & $\Gamma$(th) & $\Gamma$(exp)
& $l$ & $\Gamma$(th) & $\Gamma$(exp) \\
& & & & & & & & \\
\hline
& & & & & & & & \\
$S_{11}$   & $N(1535)$ & $244$ & & & & D &  $23$ & $1 \pm 1$ \\
$S_{11}$   & $N(1650)$ & $345$ & & & & D &  $24$ & $7 \pm 5$ \\
$P_{13}$   & $N(1720)$ & $402$ & P &   $1$ & $102 \pm 89$
                               & F &  $10$ & $$ \\
$D_{13}$   & $N(1520)$ & $230$ & S &   $3$ & $ 10 \pm 4$
                               & D &   $9$ & $ 15 \pm 3$ \\
$D_{13}$   & $N(1700)$ & $386$ & S & $111$ & $  2 \pm 4$
                               & D & $114$ & $ 14 \pm 26$ \\
$D_{15}$   & $N(1675)$ & $366$ & & & & D & $123$ & $ 88 \pm 14$ \\
$F_{15}$   & $N(1680)$ & $370$ & P &   $2$ & $ 13 \pm 5$
                             & F &   $3$ & $  1 \pm 1$ \\
$G_{17}$   & $N(2190)$ & $740$ & D &  $13$ & & G & $12$ & \\
$G_{19}$   & $N(2250)$ & $780$ &   &       & & G & $40$ & \\
$H_{19}$   & $N(2220)$ & $760$ & F &   $3$ & & H &  $3$ & \\
$I_{1,11}$ & $N(2600)$ & $1003$ & G & $4$ & & I & $3$ & \\
& & & & & & & & \\
$P_{11}$ & $N(1440)$ & $147$ & & & & P & $0.1 \, \chi_1^2$ & $87 \pm 30$ \\
$P_{11}$ & $N(1710)$ & $394$ & & & & P & $70  \, \chi_2^2$ & $41 \pm 33$ \\
& & & & & & & & \\
\hline
& & & & & & & & \\
$S_{31}$ & $\Delta(1620)$ & $320$ & & & & D & $89$ & $ 67 \pm 26$ \\
$P_{31}$ & $\Delta(1910)$ & $546$ & P &  $4$ & & & & \\
$P_{33}$ & $\Delta(1920)$ & $553$ & P & $15$ &
                                  & F & $14$ & \\
$D_{33}$ & $\Delta(1700)$ & $386$ & S & $55$ & $112 \pm 53$
                                  & D & $89$ & $ 12 \pm 10$ \\
$D_{35}$ & $\Delta(1930)$ & $560$ & P &  $0$ & & F & $0$ & \\
$F_{35}$ & $\Delta(1905)$ & $542$ & P & $18$ &
                                  & F & $27$ & \\
$F_{37}$ & $\Delta(1950)$ & $575$ & & & & F & $36$ & $ 77 \pm 20$ \\
$H_{3,11}$ & $\Delta(2420)$ & $890$ & & & & H & $11$ & \\
& & & & & & & & \\
$S_{31}$ & $\Delta(1900)$ & $539$ & & & & D & $\chi_1^2$ & $$ \\
$P_{33}$ & $\Delta(1600)$ & $303$ & P & $25 \, \chi_1^2$ & $180 \pm 143$
                                  & F & $0$       & $ 16 \pm 25 $ \\
& & & & & & & & \\
\end{tabular}
\end{table}

\clearpage
\begin{table}
\centering
\caption[]{\small
$N^{\ast} \rightarrow N \eta$ and $\Delta^{\ast} \rightarrow \Delta \eta$
decay widths of (3 and 4 star) nucleon and delta resonances in MeV.
For classification of the resonances see Table~\ref{npi}.
The experimental values are taken from \cite{PDG}.
The mixing angle for the $\eta$ mesons is $\theta_P=-23^{\circ}$
\cite{GIK}. $\chi_1$ and $\chi_2$ are defined in the caption of
Table~\ref{collff}.
\normalsize}
\label{ndeta}
\vspace{15pt}
\begin{tabular}{lcccc}
& & & & \\
State & Mass & $k$(MeV) & $\Gamma$(th) & $\Gamma$(exp) \\
& & & & \\
\hline
& & & & \\
$S_{11}$ & $N(1535)$ & $182$ &  $0.1$ & $74 \pm 39$ \\
$S_{11}$ & $N(1650)$ & $346$ &  $8$   & $11 \pm  6$ \\
$P_{13}$ & $N(1720)$ & $420$ &  $0.2$ & $$ \\
$D_{13}$ & $N(1520)$ & $150$ &  $0.6$ & $$ \\
$D_{13}$ & $N(1700)$ & $400$ &  $4$   & $$ \\
$D_{15}$ & $N(1675)$ & $374$ & $17$   & $$ \\
$F_{15}$ & $N(1680)$ & $379$ &  $0.5$ & $$ \\
$G_{17}$ & $N(2190)$ & $791$ & $11$   & $$ \\
$G_{19}$ & $N(2250)$ & $831$ &  $9$   & $$ \\
$H_{19}$ & $N(2220)$ & $811$ &  $0.7$ & $$ \\
$I_{1,11}$ & $N(2600)$ & $1054$ & $3$ & $$ \\
& & & & \\
$P_{11}$ & $N(1710)$ & $410$ & $17 \, \chi_2^2$ & $$ \\
& & & & \\
\hline
& & & & \\
$P_{31}$ & $\Delta(1910)$ & $322$ & $0.0$ & $$ \\
$P_{33}$ & $\Delta(1920)$ & $335$ & $0.5$ & $$ \\
$D_{35}$ & $\Delta(1930)$ & $348$ & $0$   & $$ \\
$F_{35}$ & $\Delta(1905)$ & $316$ & $1$   & $$ \\
$F_{37}$ & $\Delta(1950)$ & $372$ & $2$   & $$ \\
$H_{3,11}$ & $\Delta(2420)$ & $786$ & $2$ & $$ \\
& & & & \\
$S_{31}$ & $\Delta(1900)$ & $309$ & $3 \, \chi_1^2$ & $$ \\
& & & & \\
\end{tabular}
\end{table}

\clearpage
\begin{table}
\centering
\caption[]{\small
Strong decay widths for $\Delta^{\ast} \rightarrow N + \pi$
and $N^{\ast} \rightarrow N + \pi$ in MeV.
Experimental values are from \cite{PDG}.
\normalsize}
\label{regge}
\vspace{15pt}
\begin{tabular}{lcccccc}
& & & & & & \\
Resonance & $L$ & \multicolumn{4}{c} {$\Gamma$(th)} & $\Gamma$(exp) \\
& & Ref.~\cite{LeYaouanc} & Ref.~\cite{KI} & Ref.~\cite{CR} & Present & \\
& & & & & & \\
\hline
& & & & & & \\
$\Delta(1232)P_{33}$   & 0 & 70 & 121 & 108 & 116 & $119 \pm 5 $ \\
$\Delta(1950)F_{37}$   & 2 & 27 &  56 &  50 &  45 & $120 \pm 14$ \\
$\Delta(2420)H_{3,11}$ & 4 &  4 &     &   8 &  12 & $ 40 \pm 22$ \\
$\Delta(2950)K_{3,15}$ & 6 &  1 &     &   3 &   5 & $ 13 \pm 8 $ \\
& & & & & & \\
\hline
& & & & & & \\
$N(1520)D_{13}$   & 1 & & 85 & 74 & 115 & $67 \pm 9 $ \\
$N(2190)G_{17}$   & 3 & &    & 48 &  34 & $67 \pm 27$ \\
$N(2600)I_{1,11}$ & 5 & &    & 11 &   9 & $49 \pm 20$ \\
& & & & & & \\
\end{tabular}
\end{table}

\clearpage
\begin{table}
\centering
\caption[]{\small
Strong decay widths for the missing nucleon and delta resonances
below 2 GeV. The mixing angle for the $\eta$ mesons is
$\theta_P=-23^{\circ}$ \cite{GIK}.
\normalsize}
\label{ndmiss}
\vspace{15pt}
\begin{tabular}{lcccccc}
& & & & & & \\
State & Mass & Resonance & \multicolumn{4}{c} {$\Gamma$(MeV)} \\
& & & $N \pi$ & $\Delta \pi$ & $N \eta$ & $\Delta \eta$ \\
& & & & & & \\
\hline
& & & & & & \\
$P_{1,2J}$ & $N(1720)$
& $^{2}8_{J}[20,1^+]_{(0,0);0}$   &  0 &   0 &  0 & \\
$D_{1,2J}$ & $N(1875)$
& $^{2}8_{J}[70,2^-]_{(0,0);1}$   &  0 &   0 &  0 & \\
$P_{13}$   & $N(1875)$
& $^{2}8_{3/2}[70,2^+]_{(0,0);2}$ & 56 &  56 &  9 & \\
$F_{15}$   & $N(1875)$
& $^{2}8_{5/2}[70,2^+]_{(0,0);2}$ & 85 &  43 & 19 & \\
$S_{11}$   & $N(1972)$
& $^{4}8_{1/2}[70,2^-]_{(0,0);1}$   &  0 &   0 &  0 & \\
$D_{13}$   & $N(1972)$
& $^{4}8_{3/2}[70,2^-]_{(0,0);1}$   &  0 &   0 &  0 & \\
$D_{15}$   & $N(1972)$
& $^{4}8_{5/2}[70,2^-]_{(0,0);1}$   &  0 &   0 &  0 & \\
$G_{17}$   & $N(1972)$
& $^{4}8_{7/2}[70,2^-]_{(0,0);1}$   &  0 &   0 &  0 & \\
$P_{11}$   & $N(1972)$
& $^{4}8_{1/2}[70,2^+]_{(0,0);2}$ & 19 &  15 & 17 & \\
$P_{13}$   & $N(1972)$
& $^{4}8_{3/2}[70,2^+]_{(0,0);2}$ &  9 &  94 &  9 & \\
$F_{15}$   & $N(1972)$
& $^{4}8_{5/2}[70,2^+]_{(0,0);2}$ &  4 & 156 &  4 & \\
$F_{17}$   & $N(1972)$
& $^{4}8_{7/2}[70,2^+]_{(0,0);2}$ & 18 &  96 & 20 & \\
& & & & & & \\
$S_{11}$   & $N(1909)$ & $^{2}8_{1/2}[70,1^-]_{(1,0);1}$
& $22 \, \chi_1^2$ & $0.7 \, \chi_1^2$ & $0.5 \, \chi_1^2$ & \\
$D_{13}$   & $N(1909)$ & $^{2}8_{3/2}[70,1^-]_{(1,0);1}$
& $28 \, \chi_1^2$ & $0.6 \, \chi_1^2$ & $\chi_1^2$ & \\
$P_{13}$   & $N(1815)$ & $^{4}8_{3/2}[70,0^+]_{(0,1);0}$
& $30 \, \chi_2^2$ & $211 \, \chi_2^2$ & $24 \, \chi_2^2$ & \\
$S_{11}$   & $N(1866)$ & $^{2}8_{1/2}[56,1^-]_{(0,1);1}$
& $211 \, \chi_2^2$ & $84 \, \chi_2^2$ & $4 \, \chi_2^2$ & \\
$D_{13}$   & $N(1866)$ & $^{2}8_{3/2}[56,1^-]_{(0,1);1}$
& $271 \, \chi_2^2$ & $71 \, \chi_2^2$ & $7 \, \chi_2^2$ & \\
$P_{1,2J}$ & $N(1997)$ & $^{2}8_{J}[70,1^+]_{(0,1);0}$
& 0 & 0 & 0 & \\
$S_{11}$   & $N(1997)$ & $^{2}8_{1/2}[70,1^-]_{(0,1);1}$
& $3 \, \chi_2^2$ & $34 \, \chi_2^2$ & $5 \, \chi_2^2$ & \\
$D_{13}$   & $N(1997)$ & $^{2}8_{3/2}[70,1^-]_{(0,1);1}$
& $3 \, \chi_2^2$ & $30 \, \chi_2^2$ & $7 \, \chi_2^2$ & \\
& & & & & & \\
\hline
& & & & & & \\
$D_{33}$   & $\Delta(1945)$ & $^{2}10_{3/2}[70,2^-]_{(0,0);1}$
&  0 &   0 & &  0 \\
$P_{33}$   & $\Delta(1945)$ & $^{2}10_{3/2}[70,2^+]_{(0,0);2}$
&  9 & 105 & &  4 \\
$F_{35}$   & $\Delta(1945)$ & $^{2}10_{5/2}[70,2^+]_{(0,0);2}$
& 13 &  83 & &  2 \\
& & & & & & \\
$D_{33}$   & $\Delta(1977)$ & $^{2}10_{3/2}[70,1^-]_{(1,0);1}$
& $5 \, \chi_1^2$ & $7 \, \chi_1^2$ & & $3 \, \chi_1^2$ \\
$P_{31}$   & $\Delta(1786)$ & $^{2}10_{1/2}[70,0^+]_{(0,1);0}$
&$ 27 \, \chi_2^2$ & $172 \, \chi_2^2$ & & 0.0 \\
& & & & & & \\
\end{tabular}
\end{table}

\clearpage
\begin{figure}
\caption{Collective model of baryons.$\qquad\qquad\qquad
\qquad\qquad\qquad\qquad\qquad\qquad\qquad\qquad\qquad$}
\label{geometry}
\end{figure}

\begin{figure}
\caption{
Elementary meson emission.$\qquad\qquad
\qquad\qquad\qquad\qquad\qquad\qquad\qquad\qquad\qquad\qquad$}
\label{qqM}
\end{figure}

\begin{figure}
\caption{
Strong decay widths for $N^{\ast} \rightarrow N + \pi$ decays
of negative parity resonances with $L^P=1^-$.
The theoretical values are in parenthesis. All values in MeV.}
\label{nwidths}
\end{figure}

\begin{figure}
\caption{
Strong decay widths for $\Delta^{\ast} \rightarrow N + \pi$ and
$\Delta^{\ast} \rightarrow \Delta + \pi$ decays
of positive parity resonances with $L^P=2^+$ and negative parity
resonances with $L^P=1^-$.
Notation as in Fig.~\protect\ref{nwidths}.}
\label{dwidths}
\end{figure}

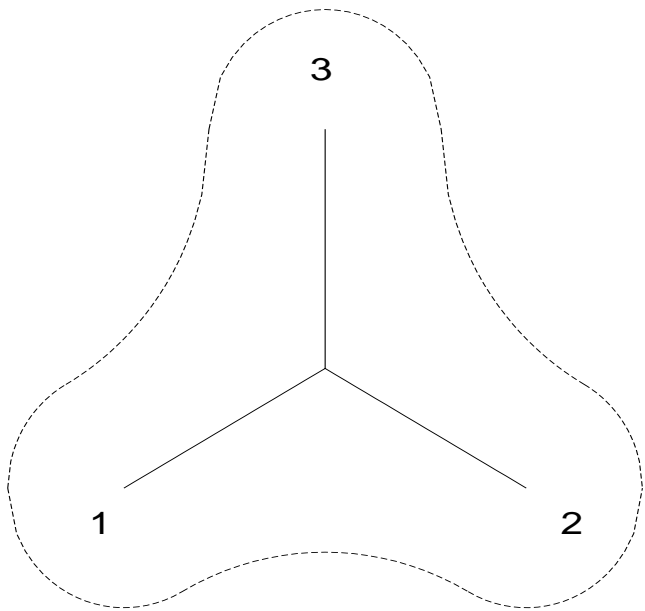
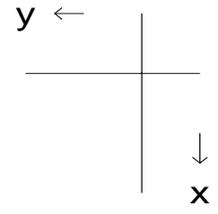
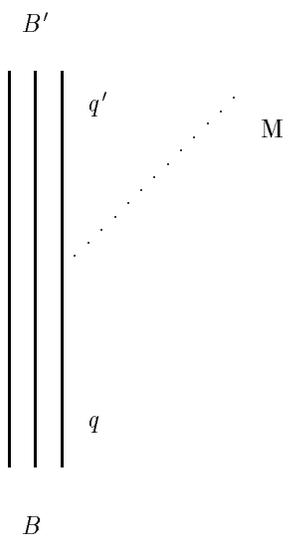
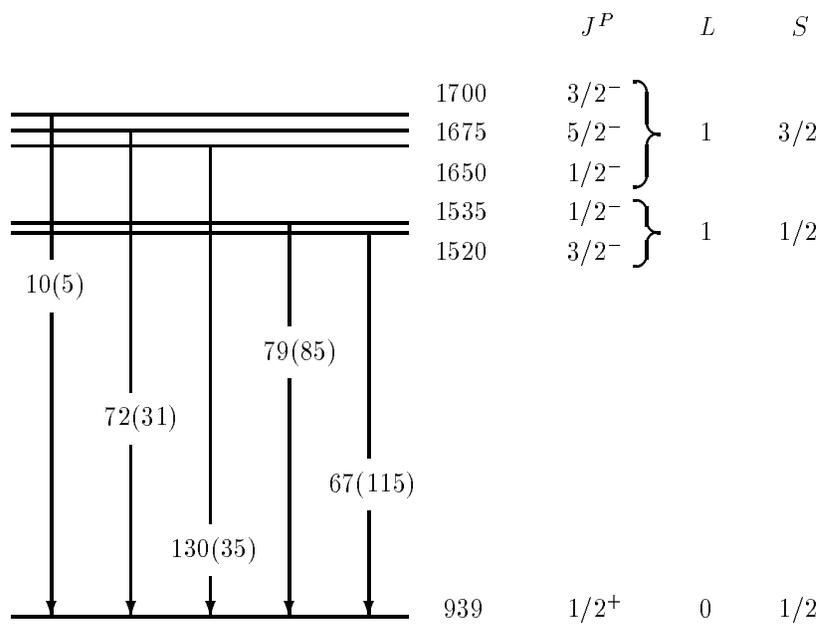
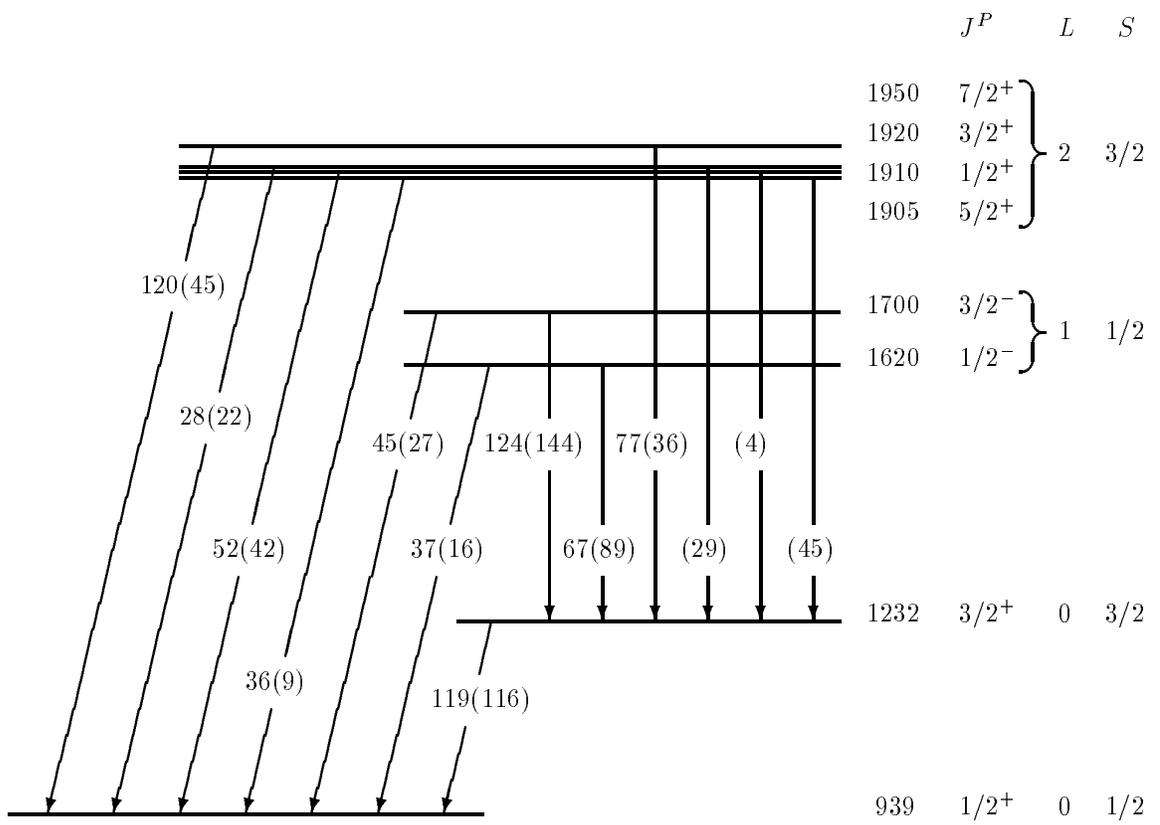
\end{document}